\documentclass[11pt]{article}

\usepackage{amsmath,amsfonts,amssymb,amsthm,bbm,stmaryrd}
\usepackage[dvips]{graphicx}
\usepackage{latexsym}
\usepackage{psfrag}
\usepackage[latin1]{inputenc}
\usepackage{hyperref}
\usepackage{color}
\usepackage{enumerate}
\numberwithin{equation}{section}
\usepackage{float}
\usepackage{cite}

\newcommand{\be}{\begin{equation}}
\newcommand{\ee}{\end{equation}}
\newcommand{\barray}{\begin{array}}
\newcommand{\earray}{\end{array}}
\newcommand{\bea}{\begin{eqnarray}}
\newcommand{\eea}{\end{eqnarray}}
\newcommand{\bs}{\begin{subequations}}
\newcommand{\es}{\end{subequations}}

\newcommand{\bit}{\begin{itemize}}
\newcommand{\eit}{\end{itemize}}
\newcommand{\bd}{\begin{description}}
\newcommand{\ed}{\end{description}}

\def\nn{\nonumber}

\newcommand{\p}{\partial}

  \newcommand{\g}{\gamma}  
  \newcommand{\eps}{\epsilon}

\let\m=\mu    \let\n=\nu

\def\cN{{\cal N}}

\newcommand{\pbi}[1]{\underset{\leftarrow}{#1}}

\begin{document}

\title{\bf A note on the physical process first law of black hole mechanics}

\author{\Large{Antoine Rignon-Bret}
\smallskip \\ 
\small{\it{Aix Marseille Univ., Univ. de Toulon, CNRS, CPT, UMR 7332, 13288 Marseille, France}} }
\date{\today}

\maketitle

\begin{abstract}
    I give a simple proof of the physical process first law of black hole thermodynamics including charged black holes, in which all perturbations are computed on the horizon.  
\end{abstract}

\section{Introduction}

One of the most mysterious and promising topics in modern physics is the deep connection between the laws of black hole mechanics and the fundamental principles of thermodynamics. Originally stated in the seminal paper of Bardeen, Carter and Hawking \cite{bardeen1973four}, the analogy between the laws of black hole mechanics and thermodynamics has been taken more seriously after the deep insights of Bekenstein about black hole entropy \cite{bekenstein2020black}, the introduction of the generalized second law \cite{Bekenstein:1974ax}, and the discovery by Hawking that black holes radiate \cite{Hawking:1974rv, hawking1975particle}. The first law of black hole thermodynamics is very general and robust. It can be extended to the case of charged black holes  using the Hamiltonian formalism \cite{Sudarsky:1992ty}, to any covariant theory of gravity using the Lagrangian formalism \cite{wald1993black, iyer1994some} including cases where arbitrary fields with internal degrees of freedom are present \cite{prabhu2017first}. In general, it relates the variations of asymptotic charges, as the ADM mass $M$ and angular momentum $J$ and the electric charge $Q$ to the variation of the black hole area $A$. This is a relation between phase space variables, on the same footing as the usual first law of black hole thermodynamics.

\bigskip

However, while usual thermodynamic relations deal with ideal transitions between equilibrium states characterized by perfectly stationary solutions, it is sometimes instructive to consider realistic transitions bringing one stationary state into another one. Indeed, it is a priori non trivial that a new equilibrium  state is reached after introducing some perturbations into the system. Equivalently, in black holes mechanics, we are interested in the evolution of the black hole if some piece of matter is thrown into it. If we assume that the black hole settles down to a new stationary state after some small energy, angular momentum and electric charge fluxes have crossed the horizon, it would be consistent to recover an identity similar to the equilibrium state version of the first law. This version of the first law of black hole thermodynamics is called the Physical Process First Law (PPFL) and its relation to the equilibrium state version of the first law is subtle \cite{mishra2018physical, sarkar2019black}. The problem of energy and angular momentum perturbations to the Kerr black hole has been studied first by Hawking and Hartle \cite{hawking1972energy}. The usual PPFL has been derived by Wald in \cite{wald1994quantum} for uncharged black holes and extended to the charged case by Gao and Wald in \cite{gao2001physical}. 

\bigskip

However, while the mass and angular momentum variations are calculated directly on the horizon for the perturbed Kerr black hole \cite{wald1994quantum}, it is not the case for the derivation in the charged case in \cite{gao2001physical}. Indeed, this derivation has the inconvenience of relying on the covariant phase space techniques. In particular, the black hole mass and angular momentum variations are computed at spatial infinity, as the ADM mass and angular momentum. It might be a bit unexpected for a physical process first law to involve variations of asymptotic quantities rather than the mass or angular momentum of the matter fields crossing the black hole horizon. In fact, Gao and Wald wrote \footnote{see footnote 5 of their paper} that the ADM charges $M$ and $J$ should hold as the definition for the final mass and angular momentum of the black hole even if not all the matter fell into it \cite{gao2001physical}, but they did not explicitly show that this is indeed the case. As they work with the ADM masses and angular momentum, they cannot distinguish the mass and angular momentum of the black hole from the one of the matter outside. In contrast, the derivation presented below is similar to the original derivation of the PPFL from Wald \cite{wald1994quantum} for the uncharged black hole, where the mass and angular momentum variations are given locally and are not the perturbed ADM Hamiltonians at infinity. We compute everything on the horizon and do not need to introduce the covariant phase space formalism.

\section{Derivation of the PPFL}
\label{section3}

\vspace{0.3 cm}
\begin{figure}[t]
\begin{center}
\includegraphics[scale=1.4]{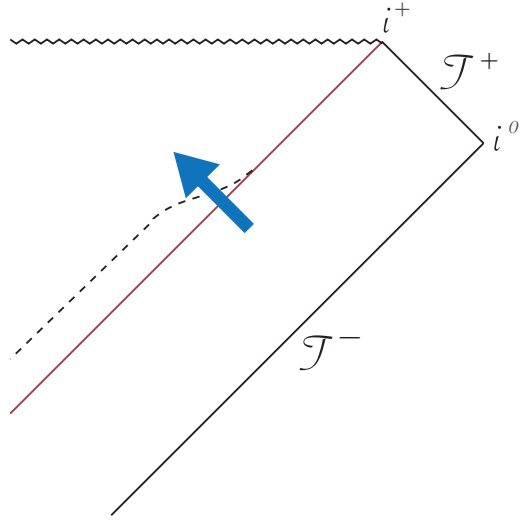}
\caption{Penrose diagramm sketching a piece of matter falling into a black hole (blue arrow). The null event horizon is represented in red, and is defined as the boundary of the past of future null infinity, as usual. The apparent horizon is a dashed black line.}
\label{Figurepenrose}
\end{center}
\end{figure}
Let us consider a stationary black hole, with an event horizon $\cN$ parameterized by some null affine parameter $v$ vanishing on the black hole bifurcation surface $v = 0$. This eternal black hole has mass $M$, angular momentum $J$ and charge $Q$. The Killing vector $\xi$ generating the null geodesics of the event horizon is defined as

\be
\xi^\m = (\frac{\p}{\p t})^\m + \Omega_H (\frac{\p}{\p \phi})^\m \stackrel{\cN}= \kappa v n^\m
\label{Killingvector}
\ee
where $\frac{\p}{\p t}$ is a Killing vector which is timelike and normalized to one at infinity, $\frac{\p}{\p \phi}$ is an axisymmetric Killing field generating closed orbits of length $2 \pi$, $n^\m = (\frac{\p}{\p v})^\m$ the null normal with vanishing unaffinity, $\Omega_H$ the horizon's angular velocity, and $\kappa$ being the unaffinity of $\xi$ on $\cN$, which turns out to be the surface gravity of the stationary black hole. As the black hole is charged, there is some electromagnetic potential $A_\mu$ alongside the metric $g_{\mu \nu}$, which are the Kerr-Newman solutions of the Einstein-Maxwell equations. We use the same gauge for the Kerr-Newman solution as in \cite{wald2010general}. This is the stationary gauge for which the electromagnetic potential satisfies the stationarity requirement $\pounds_\xi A_\mu = 0$, in addition to $\pounds_\xi g_{\m \n} = 0$. Furthermore, in this gauge, $A_\m$ vanishes at spatial infinity. It is also the same gauge as the one used in \cite{gao2001physical, gao2003first}.

\vspace{0.3 cm}   

Now, let us assume that the stationary black hole is perturbed by some small amount of matter propagating on spacetime and crossing the black hole horizon at some point, such that at very late times ($v$ going to infinity), the black hole has settled down into a stationary state (see Figure.\ref{Figurepenrose}). Hence, the event horizon $\cN$ becomes a Killing horizon a large amount of time after the incoming matter crossed $\cN$, and during the whole process, the spacetime is a slightly perturbed Kerr-Newman spacetime. If we consider some (charged) perturbed matter field of order $\eps$, its corresponding stress energy tensor is of order $\eps^2$ and so the perturbation to the stationary background metric is of order $\eps^2$ at most. Furthermore, the charge current $J^\m$ is also of order $\eps^2$ \footnote{For instance, in Dirac fields electrodynamics, a Dirac field would be of order $\eps$ and the associated current will be $J^\m = i \bar{\psi} \gamma^\m \psi = O(\eps^2)$. Same for scalar electrodynamics, if the scalar field $\phi$ is of order $\eps$, the current is given by $J^\m = i (\bar{\phi} \p^\mu \phi - \p^\m \bar{\phi} \phi) = O(\eps^2).$}. Hence, we have

\begin{align}
    \phi &= O(\eps) \nn \\
    T_{\m \n}^{matter} &= O(\eps^2) \nn \\
    J^\m &= O(\eps^2)
    \label{perturbedfields}
\end{align}
from which it implies by using the equations of motion

\begin{align}
    \pounds_\xi g_{\m \n} &= O(\eps^2) \nn \\
    \pounds_\xi A_\mu &= O(\eps^2)
    \label{perturbedkilling}
\end{align}
as $\xi$ being of course the background Killing, and so that $\xi^\m \stackrel{\cN}= \kappa v n^\m$ is still null everywhere on the (dynamical) event horizon. The Raychaudhuri equation on the perturbed event horizon $\cN$ for the affine normal vector $n$ reads

\be
    \frac{d \theta_n}{dv} = - \theta_n^2 -\sigma_{n, \m \n} \sigma_n^{\m \n} - T_{\m \n} n^\m n^\n
    \label{Rayeq}
\ee
where $\theta_n$ is the expansion of $n^\m$, $\sigma_{n, \m \n}$ the shear of $n^\m$, and there is no twist $\omega_{\m \n} $ because the null vector $n$ is hypersurface orthogonal. The expansion and the shear are of order $\eps^2$ from \eqref{perturbedkilling}, so if we keep only the leading order terms in \eqref{Rayeq} we get 

\be
    \frac{d \theta_n}{dv} = - T_{\m \n} n^\m n^\n + O(\eps^4)
    \label{Rayeq2}
\ee
Now, we can multiply both sides of the equation \eqref{Rayeq} by $\kappa v$ and integrate by part the left hand side. The boundary terms vanish as we integrate between the bifurcation surface of the background spacetime located at $v = 0$ and infinity, where $v \theta_n \underset{v\rightarrow+\infty}{\rightarrow} 0$. Hence we get from \eqref{Rayeq2}

\be
\frac{\kappa}{8 \pi} \Delta A = \int_\cN T_{\m \n} \xi^\m n^\n \eps_N + O(\eps^4)
\label{linrayeq}
\ee
Terms on both sides of this equation are of order $\eps^2$. We can notice that the null energy conditions are satisfied, the RHS is positive, and so is the LHS, in agreement with Hawking's area theorem \cite{Hawking:1971tu}. Furthermore, we would like to write the RHS of \ref{linrayeq} as a linear combination of the mass, angular momentum and charge of the matter perturbation crossing the null horizon. First, consider the case where the black hole is not charged. Therefore, $T_{\m \n} = T_{\m \n}^{matter}$ and if $\xi$ is a background Killing vector \ref{Killingvector}, we have from \ref{perturbedfields}, \ref{perturbedkilling}

\be
\nabla_\m (T_\n^\m \xi^\n) = O(\eps^4)
\label{current conservation}
\ee
as $T_{\mu \nu}$ is symmetric and divergence free if the equations of motion are imposed. Furthermore, as $\frac{\p}{\p t}$ and $\frac{\p}{\p \phi}$ are also Killing vectors of the background metric, we also have $\nabla_\m \big( T_\n^\m (\frac{\p}{\p t})^\n \big) = O(\eps^4)$ and $\nabla_\m \big( T_\n^\m (\frac{\p}{\p \phi})^\n \big) = O(\eps^4)$. Hence, the currents 

\be
j_t^\m = T_\n^\m (\frac{\p}{\p t})^\n 
\ee
and 
\be
j_\phi^\m = -T_\n^\m (\frac{\p}{\p \phi})^\n
\ee
are conserved up to $\eps^4$ terms, so they are conserved at first order in perturbation of the metric and the stress energy tensor, i.e at order $\eps^2$. The physical meaning of these quantities is clear when we look at these currents at infinity, and so we identify $j_t$ as the energy current and $j_\phi$ as the angular momentum current. Hence, to make sense of \ref{linrayeq}, we can apply Gauss's theorem between Cauchy surfaces $\Sigma_1$ and $\Sigma_2$ intersecting the null event horizon in two cross sections $S_1$ and $S_2$ respectively. In order to relate these considerations to \eqref{linrayeq}, we suppose that $S_1$ is the bifurcation surface located at $v = 0$ and that $S_2$ is in the very far future ($v \longrightarrow + \infty$) such that the perturbations on the horizon vanish well before $S_2$. Hence, \ref{linrayeq} becomes

\be
    \frac{\kappa}{8 \pi} \Delta A = \Delta M - \Omega_H \Delta J + O(\eps^4)
    \label{PPFL}
\ee
However, if the black hole is charged, the relation \eqref{current conservation} is not satisfied. Indeed, we have

\be
T_{\m \n} = T_{\m \n}^{EM} + T_{\m \n}^{matter}
\label{totalstressenergy}
\ee
with \cite{blaschke2016energy}

\be
T^{EM}_{\m \n} = \frac{1}{4}(F_{\m \rho} F_\nu^\rho - \frac{1}{4} g_{\m \n} F^{\rho \sigma} F_{\rho \sigma})
\label{emtensor}
\ee
which is of order $1$. Hence

\be
    \nabla_\m (T_\n^\m \xi^\n) = T_{\m \n}^{EM} \nabla^{(\mu} \xi^{\nu)} + O(\eps^4) = O(\eps^2)
    \label{divT}
\ee
is of order $\eps^2$ and not of order $\eps^4$ as in a non-charged case \eqref{current conservation}. Therefore, we cannot apply Gauss theorem, because the current is not divergence free at order $\eps^2$. Thus we have to proceed a bit differently. It can be proven from \eqref{emtensor} and using Bianchi identity that

\be
\nabla^\m T^{EM}_{\m \n} = J^\m F_{\m \n} 
\label{step1}
\ee
which is not conserved of course if the electromagnetic field is coupled to matter. However, on-shell, the full stress energy tensor $T_{\mu \nu}$ is still divergence free on-shell and then we get from \eqref{step1}

\be
   - \nabla^\m T^{matter}_{\m \n} = J^\m F_{\m \n} 
   \label{step2}
\ee
Then we contract both sides of the equation by $\xi^\n$, and using $F_{\m \n} = \nabla_\m A_\n - \nabla_\n A_\m$, the charge conservation $\nabla_\m J^\m = 0$ and the identity $\pounds_\xi A_\m = \xi^\n \nabla_\n A_\m + A_\n \nabla_\m \xi^n$, we get from \eqref{step2}

\be
    \nabla^\m(T^{matter}_{\m \n} \xi^\n + J_\m A_\n \xi^\n) = T_{\m \n}^{matter} \nabla^\m \xi^\n + J^\m \pounds_\xi A_\m
    \label{step3}
\ee
Furthermore, from \eqref{perturbedfields} and \eqref{perturbedkilling}, we can conclude that

\be
    \nabla^\m(T^{matter}_{\m \n} \xi^\n + J_\m A_\n \xi^\n) = O(\eps^4)
    \label{step4}
\ee
Hence, the current 

\begin{equation}
    j^\m = (T^{matter})_\n^\m \xi^\n + J^\m A_\n \xi^\n
\end{equation}
is conserved at order $\eps^4$, and so are the energy current 

\be
    j_t^\m = (T^{matter})_\n^\m (\frac{\p}{\p t})^\n + J^\m A_\n (\frac{\p}{\p t})^\n
\ee
and angular momentum current \footnote{It can be seen that the current $j_t$ and $j_\phi$ correspond to the energy and angular momentum currents respectively when evaluated at infinity, because $ \lim_{r\rightarrow\infty} A_\mu = 0$ in the stationary gauge.} 

\be
    j_\phi^\m = -\big[ (T^{matter})_\n^\m (\frac{\p}{\p \phi})^\n + J^\m A_\n (\frac{\p}{\p \phi})^\n \big]
\ee
Again, the physical meaning of these currents is clear if we look them on the asymptotically flat spacetime. Indeed, at infinity, the background electromagnetic potential $A_\m$ vanishes in the stationary gauge we chose, and so at infinity the energy and angular momentum are given only by the stress energy tensor of the charged matter, as expected in Minkowski'spacetime. Therefore, we should make appear these conserved currents in \eqref{linrayeq}, in order to apply Gauss theorem between the Cauchy slices $\Sigma_1$ and $\Sigma_2$ the piece of the event horizon $\cN$ comprised between $S_1$ and $S_2$. Hence, we write \eqref{linrayeq} as 

\begin{align}
\frac{\kappa}{8 \pi} \Delta A &= \int_\cN (T^{matter}_{\m \n} \xi^\m n^\n + T^{EM}_{\m \n} \xi^\m n^\n) \eps_N  + O(\eps^4) \\
&= \int_\cN [(T^{matter}_{\m \n} + J_\n A_\m) \xi^\m n^\n - J^\nu A_\mu \xi^\mu n_\n \big) \eps_\cN + \int_\cN T^{EM}_{\m \n} \xi^\m n^\n \eps_\cN]
\label{linrayeqcharge}
\end{align}
Let us start by looking at the first integral. The first term in the integral is our conserved current. Furthermore, we can see that $J^\n n_\n \eps_\cN = - dQ$ is the infinitesimal electric charge crossing the horizon, and $\Phi_H = - \xi^\m A_\m$ is the electrostatic potential of the horizon, constant on a non-expanding horizon. Indeed, we have

\be
 \pounds_\xi A = d i_\xi A + i_\xi F
 \label{liexiA}
\ee
as $F = dA$. If we take the pullback of \eqref{liexiA} on the non-expanding null horizon $\cN$, we get $\pbi{di_\xi A} = d\pbi{i_\xi A} = 0$ as $\xi$ is tangent to $\cN$. Hence, $d\Phi_{H} = O(\eps^2)$ on the perturbed horizon, and from \eqref{step4} the first integral of \eqref{linrayeqcharge} can be written as

\be
    \int_\cN [(T^{matter}_{\m \n} + J_\n A_\m) \xi^\m n^\n - J^\nu A_\mu \xi^\mu n_\n] \eps_\cN = \Delta M - \Omega_H \Delta J - \Phi_{H} \Delta Q + O(\eps^4)
    \label{firstint}
\ee
Now, we should look at the second integral of the RHS in \eqref{linrayeqcharge}. On a non-expanding horizon, we have $T_{\m \n}^{EM} \xi^\m n^\n = 0$ from the Raychaudhuri equation \eqref{Rayeq}, ie.  $T_{\m \n}^{EM} \xi^\m n^\n = \kappa v F_{\m \n}\xi^\n F_{\rho} n^\rho = 0$. It means that the vector $F^{\mu \nu} n_\n$ is null. Furthermore, $F_{\mu \nu} n^\m n^\nu = 0$ by antisymmetry of $F$, so the vector $F^{\mu \nu} n_\n$ is tangent to $\cN$. As it is tangent to $\cN$ and null, it means that it is proportional to $n^\m$, and we can write $F^\m_\n n^\n = \alpha n^\m$. From this, we deduce that on the dynamical event horizon we have 

\be
    F^{\m \n} n_\n = \alpha n^\m + \eps^2 \gamma^\m
    \label{Fhorizon}
\ee
as the first order of perturbation in $F$ is $\eps^2$ (as we use the electromagnetic equations of motion and we know that $J^\m$ is of order $\eps^2$) for some vector $\g^\m$. Then, by antisymmetry of $F$ 

\begin{align}
    0 &= F^{\m \n} n_\m n_\n = (\alpha n^\m + \eps^2 \gamma^\m) n_\m = \eps^2 \gamma^\m n_\m
    \label{gammaN}
\end{align}
so $\gamma^\m n_\m = 0$ and $\gamma^\m$ is tangent to $\cN$. Therefore, on the dynamical null horizon, we get

\begin{align}
    T^{EM}_{\m \n} \xi^\m n^\n &= \kappa v (\alpha n^\m + \eps^2 \gamma^\m)(\alpha n_\m + \eps^2 \gamma_\m) \nn \\
    &= \kappa v \gamma^\m \gamma_\m \eps^4 = O(\eps^4)
    \label{secondint}
\end{align}
Hence, from \eqref{linrayeq}, \eqref{firstint} and \eqref{secondint}, we deduce the PPFL for charged black holes

\be
    \frac{\kappa}{8 \pi} \Delta A = \Delta M - \Omega_H \Delta J - \Phi_H \Delta Q + O(\eps^4)
    \label{finalstep}
\ee
where here, $\Delta M$, $\Delta J$ and $\Delta Q$ correspond to the energy, angular momentum and charge of the piece of matter flowing through the event horizon. We should notice that the term $T_{\m \n} \xi^\m n^\n$, usually associated to an energy flow, is more related to the entropy variation than to the energy variation. Furthermore, though we restricted ourselves to electromagnetism, there should not be any difficulties to extend the derivation above to arbitrary Yang-Mills fields and charged matter.

\section*{Acknowledgements}

I would like to thank Alejandro Perez, Simone Speziale and Robert Wald for helpful discussions.

\bibliographystyle{unsrt}
\bibliography{bibliographe.bib}
\end{document}